\documentclass[12pt,a4paper]{article}
\usepackage{multicol}
\usepackage{graphicx}

\begin{document}
\baselineskip=18pt

\author{A.Loskutov, I.A.Istomin, K.M.Kuzanyan and O.L.Kotlyarov\\
{\small Physics Faculty, Moscow State University,
Moscow 119899 Russia}}

\title{Testing and forecasting the time series of the
solar activity by singular spectrum analysis}

\date{ }

\maketitle

\begin{abstract}

To study and forecast the solar activity data a quite perspective
method of singular spectrum analysis (SSA) is proposed. As known,
data of the solar activity are usually presented via the Wolf numbers
associated with the effective amount of the sunspots. The advantages
and disadvantages of SSA are described by its application to the
series of the Wolf numbers. It is shown that the SSA method provides
a sufficiently high reliability in the description of the 11-year solar
cycle. Moreover, this method is appropriate for revealing more long
cycles and forecasting the further solar activity during one and a half
of 11-year cycle.

\end{abstract}

\vspace{0.3cm}

\begin{multicols}{2}

\section{Introduction}

It has long been observed that solar activity depends on a
number of spots visible on its disk. During about 11 years
which is called a solar cycle, this number varies over a wide
range. Accompanying this process changes in the structure
of magnetic fields of the Sun affect the Earth climate and have
a probable connection with the natural catastrophes. In addition,
intensity of the solar radiation (the frequency of the Sun flares,
etc.) apparently exerts an influence on all areas of human life
including socially-historical activity. Thus, in view of the quite large
significance of the magnetic activity of the Sun its prediction is a
subject of much current interest.

At the present time, to describe the dynamics of the solar activity
many approaches are used. Among them the Wolf number associated
with the effective amount of the sunspots is much more convenience
method. The dynamics of the Wolf numbers has more or less
quasiperiodic nature but, owing to the facts that models of this
process do not take into account many essential factors of the
solar magnetic activity, its prediction is difficult. It is known that
during last 250 years the cycle period of the sunspots changes its
value not more than $20\%$. At the same time, the amplitude
(i.e. the averaged number of sunspots) has varied over an order
or even more. Detailed models of the solar processes do not describe
such variations.

In the last few years sufficiently many methods devoted to the
prediction and reconstruction (to the past) the dynamics of the Wolf
numbers have been proposed (see, e.g., \cite{Scha, Nagovit, Wilhar,
Hos, Hawr} and references cited therein). It should be noted
however, that these methods have certain limitations and demerits.
That is the reason why prediction of the sunspot dynamics based on
the data of observations only (i.e. without a modelling of the process)
is a quite perspective approach. In this way time-series analysis
(see \cite{Afreiman, Cas, Lomi, Ruelle, Sayoca, Malinpot})
can give an essential assistance. But in this case there are many
problems related to the fact that the Wolf series is not strictly
deterministic system, it does not have clearly defined dimension
\cite{Lorca1, Lorca2}, and its length is not so large.

In the present paper, for investigations of the Wolf time series
a sufficiently efficient method of singular spectrum analysis is
proposed. It is shown that this method provides a quite high
reliability in the description of amplitudes of the 11-year solar
cycle. Moreover, it is appropriate for revealing more long cycles
and forecasting the further solar activity during one and a half of
11-year cycle, i.e. the nearest 16 years.

\section{Singular spectrum \\ analysis}

The method of singular spectrum analysis (SSA) \cite{Broomki1,
Broomki2, Brojo, Vayighil, Dazhi}, which is used in the present article
allows us the following.

\begin{itemize}

\item

Recognise certain components in the time series which have been
obtained from the observable at regular intervals;

\item
Find periodicities that are not known in advance;

\item
On the basis of the chosen components to smooth out the initial data;

\item
Extract components with the known period;

\item
Predict the further evolution of the observed dependence.

\end{itemize}

The method of SSA is a reasonably new one, but now it is quite clear
that it is sufficiently competitive with numerous smoothing methods
\cite{Perwal, Theelgfa, Kagl}. Moreover, in certain cases forecasting
the system evolution on its basis gives much more reliable results in
comparison with the other known algorithms (see \cite{Cas, Dazhi,
Debge, Murray, Cahofahe, Kepghil}  and references therein).

The SSA method is based on the passage from the investigation of
an initial linear series $(x_i)_{i=1}^N$ to the analysis of a
many-dimensional series consisted of components of some
length which contain except for the value $x_i$, certain quantities
$x_{i-j}$, $j=1,\ldots,\tau$, at the previous instants of time.

Let us describe the central steps of the application of SSA to a series
$(x_i)_{i=1}^N$.

(1) On the first step, a one-dimensional series is transformed into
a many-dimensional one. For such a transformation it is necessary
to take a certain number of delays $\tau\leq [N/2+1]$, where $[\cdot]$
is an integer part, and represent the initial $\tau$ values of the
sequence as the first column of some matrix $X$. For the second
column, the values of the sequence from $x_2$ to $x_{\tau+1}$ are
chosen. Thus, the last $\tau$ elements of the sequence
$x_n,\ldots,x_N$ correspond to the last column with the number
$n=N-\tau+1$. Therefore the transformed series has the following
matrix form:
$$
X=\left(
\begin{array}{ccccccc}
x_1 & x_2 & \ldots & x_\tau & \ldots & x_n \\
x_2 & x_3 & \ldots & x_{\tau+1} & \ldots & x_{n+1} \\
x_3 & x_4 & \ldots & x_{\tau+2} & \ldots & x_{n+2} \\
\vdots & \vdots & \ddots & \vdots & \ddots & \vdots \\
x_{\tau} & x_{\tau+1} & \ldots & x_{2\tau-1} &
\ldots & x_N \\
\end{array}
\right).
$$
It is obvious that for the constructed matrix the expression
$||x_{ij}||=x_{i+j-1}$ holds. In general, the matrix $X$
is a rectangular one. But, in a limit case, i.e. at $\tau=N/2$
and an even $N$, it degenerates into a square matrix.

(2) After this transformation, for the matrix $X$ the corresponding
covariance matrix
$$
C=\frac{1}{n} XX^T
$$
should be obtained.

(3) Now it is necessary to find eigenvalues and eigenvectors
of the matrix $C$. For this, the matrix $C$ should be factored
as follows: $C=V\Lambda V^T$, where
$\Lambda=\left(
\begin{array}{cccc}
\lambda_1 & 0 & \ldots & 0 \\
0 & \lambda_2 & \ldots & 0 \\
\vdots & \vdots & \ddots & \vdots\\
0 & 0 & \ldots & \lambda_\tau\\
\end{array}
\right)$ is the diagonal matrix of eigenvalues,
$
V=\Bigl(V^1,V^2,\ldots,V^\tau\Bigr)=
\left(
\begin{array}{cccc}
v_1^{1} & v_1^{2} & \ldots & v_1^{\tau}\\
v_2^{1} & v_2^{2} & \ldots & v_2^{\tau}\\
\vdots & \vdots & \ddots & \vdots\\
v_\tau^{1} & v_\tau^{2} & \ldots & v_\tau^{\tau}\\
\end{array}
\right)
$ is the orthogonal matrix of eigenvectors of the matrix $C$.
It is clear that $\Lambda=VCV^T$, $\sum_{i=1}^\tau\lambda_i=\tau$
and $\det C=\prod_{i=1}^\tau\lambda_i$.

(4) For the next step, the matrix $V$ of eigenvectors should be
presented as a matrix of the conversion to the principal components,
$Y=V^T X=(Y_1,Y_2,\ldots,Y_{\tau})$, of the initial series.
Here $Y_i$, $i=1,2,\ldots,\tau$, are rows of the length $n$. Therewith,
the eigenvalues $\lambda_1,\lambda_2,\ldots,\lambda_\tau$ can be
considered as a certain contribution of the principal components
to a general information content of the series $(x_i)_{i=1}^N$.
Then, by means of these principal components it is possible to
reconstruct the initial matrix $X$:
$$
X=\left(V^1,V^2,\ldots,V^\tau\right)\left(
\begin{array}{c}
Y_1^T \\
Y_2^T \\
\vdots \\
Y_\tau^T \\
\end{array}
\right)=\sum_{i=1}^{\tau}V^iY_i^T.
$$
In turn, by the matrix $X$ one can reconstruct the time series
$(x_i)_{i=1}^N$. It should be noted that for the reconstruction
not all principal components $Y_1,Y_2,\ldots,Y_{\tau}$ are usually
applied. Only a part of them can be involved. This depends on the
goal which we pursue and the informative content of the used
components (see \cite{Broomki1, Broomki2, Brojo, Vayighil}).
This means that each vector-row $Y_i$ can be considered as a
result of some projection of a $\tau$-dimensional totality on a direction
corresponding to the eigenvector $V^i$. Thus, the series is presented
via a set of $\tau$ components $Y_i$. Therewith, the weight of the
component $Y_i$ in the initial sequence $(x_i)_{i=1}^N$ can be
defined by the corresponding eigenvalue $\lambda_i$ which is,
in turn, the eigenvalue of the eigenvector $V^i$.

Each $i$-th eigenvector includes $\tau$ components,
$V^i=\left(
\begin{array}{c}
v^i_1\\
v^i_2\\
\vdots \\
v^i_\tau\\
\end{array}
\right)$.
Let us construct a dependence of the component values
$v^i_k$, $k=1,2,\ldots,\tau$, as a function of their number:
$v^i=v^i(k)$. Then using the orthogonality property of eigenvectors,
the further analysis of the sequence $(x_i)^N_{i=1}$ can be
performed by means of diagrams plotted by the analogy with
the Lissajous figure. Namely, along the axes the components
$v^i_k$, $v^j_k$ are plotted in pairs. If the constructed diagrams
are similar to a circle, then the functions $v^i=v^i(k)$, $v^j=v^j(k)$
are approximated by certain periodic functions with almost coincided
amplitudes and a phase lag about a quarter of the period.

Thus, for some pairs of the eigenvectors $V^i$, $V^j$ one can find a
value which has a meaning of the period. Therefore, such a graphical
representation provides a certain estimate of the component
frequencies inherent in the initial time series $(x_i)_{i=1}^N$.

For the given parameter $\tau$ the number of all possible pairs
of principal components is $\sim\tau^2$. It is obvious that even for a
sufficiently small $\tau$ the analysis of all such pairs is a quite
laborious problem. Moreover, at a large values of $\tau$ only
a small part of diagrams has a helical form. Thus, before a
graphical analysis it would be reasonable to restrict our search.
This can be done if we arrange $V^i$ and $Y_i$ in order of
decreasing their eigenvalues and consider only such pairs of
eigenvectors which have close enough eigenvalues $\lambda$.
In the diagram $\lambda=\lambda(i)$ at a quite large $\lambda$
these pairs form a decreasing (with the growth of $i$) step function.

(5) Suppose now that for the further reconstruction we have only $r$
leading components. Thus, for the reconstruction of the initial matrix
$X$ one should use $r$ leading eigenvectors $V^i$. In this case
$$
\tilde X=
\left(V^1,V^2,\ldots,V^r\right)\left(
\begin{array}{c}
Y_1\\
Y_2\\
\vdots \\
Y_r\\
\end{array}
\right)=\sum_{i=1}^{r}V^iY_i \ ,
$$
where $\tilde X$ is a reconstructed matrix with $n$ columns and
$\tau$ rows. Then the initial time series obtained from this matrix
is defined as follows:
$$
\tilde{x}_s=\left\{
\begin{array}{lr}
\displaystyle\frac{1}{s}\sum\limits_{i=1}^{s}
\tilde{x}_{i,s-i+1}\ , \ \ \ \ \ \ \ \ \ \ \ \ 1 \le s \le \tau\ ,\\
\displaystyle\frac{1}{\tau}\sum\limits_{i=1}^{\tau}
\tilde{x}_{i,s-i+1}\ , \ \ \ \ \ \ \ \ \ \ \ \ \tau \le s \le n\ ,\\
\displaystyle\frac{1}{N-s+1}\sum\limits_{i=1}^{N-s+1}
\tilde{x}_{i+s-n,n-i+1}\ ,\\
\ \ \ \ \ \ \ \ \ \ \ \ \ \ \ \ \ \ \ \ \ \ \ \ \ \ \ \ \ \ n \le s \le N\ .
\end{array}
\right.
$$
Described way of the reconstruction is said to be a SSA-smoothing
of the initial time series $(x_i)_{i=1}^N$ by the leading $r$
components.

(6) In the next stage of the SSA application, a prediction procedure
of the initial time sequence can be considered (see \cite{Kepghil,
Danilov, Ghil}). This means that the series $(x_i)_{i=1}^{N+p}$ which
is the extension of the known data $(x_i)_{i=1}^N$, is constructed. In
turn, extrapolation to $p$ points forward is reduced to the application
of $p$ times of the prediction procedure to the one point. The basic
idea of the computation of the point $x_{N+1}$ is the following.

Consider the sequence $x_1,x_2,\ldots,x_N$ and construct
a sample in the form of matrix $X$. As a basis of the surface
containing this sample one can take the chosen before vectors
$V^1,V^2,\ldots,V^r$ of the matrix $C$. Let us write the
parametric equation of this surface as
$S(P)=\sum\limits_{i=1}^{r}p_i V^{i}$, where a set of the parameters
$p_i$ corresponds to the value $S(P)$ which is a column with
$\tau$ elements. In this case the set of parameter values
$P^k=\left(p_1^k,p_2^k,\ldots,p_r^k\right)$ corresponds to
$k$-th, $k=1,2,\ldots,n$, column of the matrix $X$. Therefore,
$X^1=S\left(P^{1}\right)$, $X^2=S\left(P^{2}\right)$, $\ldots$ ,
$X^n=S\left(P^{n}\right)$.
Now, to predict the value $x_{N+1}$ it is necessary to find the
$(n+1)$-th column $X^{n+1}$ which, in turn, fits the parameters
$P^{n+1}=\left(p_1^{n+1},p_2^{n+1},\ldots,p_r^{n+1}\right)$.
Using the data $(x_i)_{i=1}^N$ these parameter values can be
obtained from the expression $S(P)=\sum\limits_{i=1}^r p_iV^i$.
Thus, the predicting column is written as follows:
$X^{N+1}=S\Bigl(P^{n+1}\Bigr)$.

Let us introduce the following designations:
$$
V_*=\left(\begin{array}{cccc}
v_1^{1} & v_1^{2} & \ldots & v_1^{r} \\
v_2^{1} & v_2^{2} & \ldots & v_2^{r} \\
\vdots & \vdots & \ddots & \vdots \\
v_{\tau-1}^{1} & v_{\tau-1}^{2} & \ldots & v_{\tau-1}^{r} \\
\end{array}\right);
$$
$$
\tilde{p}=\left(\begin{array}{c}
\tilde{p}_1^{n+1} \\
\tilde{p}_2^{n+1} \\
\vdots \\
\tilde{p}_r^{n+1} \\
\end{array}\right);
\quad
Q=\left(\begin{array}{c}
x_{N-\tau+2} \\
x_{N-\tau+3} \\
\vdots \\
x_N \\
\end{array}\right);
$$
$$
V_\tau=\left(v_\tau^{1},v_\tau^{2},\ldots,v_\tau^{r}\right).
$$
The set of the values
$\left(p_1^{n+1},p_2^{n+1},\ldots,p_r^{n+1}\right)$
is easily found as a solution of the system $V_*\tilde{P}=Q$
with respect to $\tilde P$. Thus, the final expression of the
forecasting value has the following form:
$$
x_{N+1}=\frac{V_\tau V_{*}^T Q}{1-V_\tau V^T}.
$$

In the simplest case, to predict the next values it is necessary
to change the matrix $Q$ in a corresponding way and multiply it
by the value $V_\tau V_{*}^T\Big/\left(1-V_\tau V^T\right)$. In addition,
however for each predicted point one can completely repeat the SSA
algorithm. Then the matrixes $V_\tau$ and $V_*$ will be changed.

(7) At the final stage of the SSA application one should dwell on
the choice of the main parameter --- the number of delays $\tau$
which are used for the many-dimensional sample $X$. As in the case
of selection of the principal components this value essentially depends
on the investigated problem.

Consider the smoothing procedure of a time series by the SSA
method. In this case, as noted above, selection of a principal
component is the filtration of the time series with the transition filtering
function in the form of the eigenvector of this principal component. If
the delay value $\tau$ is the greater, the greater the number of
parallel filters, and a bandwidth of each of them is more narrow.
Thus, for a large enough $\tau$ we have a sufficiently efficient
smoothing of the time series.

If it is necessary to define unknown (hidden) periods in the observed
sequence then one should take the value of $\tau$ as large as
possible. Next, after omitting close to zero eigenvalues, the delay
value should be shortened.

Suppose now that we should select only the one known periodic
component. In this case it is necessary to choose the delay
$\tau$ which is equal to the required period.

Finally, let us consider the problem of some extension of an
observed sequence to a given value, i.e. the problem of forecasting
the evolution of the process under investigation. Then one should
take a maximum allowable value of the delay $\tau$ and thereafter
select the number $r$.

\section{Forecasting the solar activity by SSA}

To estimate of the practicality of the SSA method it is necessary to
use the time series of a natural origin. In the present article, as such
a time series the sequence of the Wolf numbers characterised
in a certain way the solar activity was chosen. Taken alone, the Wolf
numbers defined by the visible sunspots cannot give a quantitative
information related to the solar activity. However there is a large
enough correlation between the Wolf numbers and the $F10.3$
emission. That is the reason why investigations of the relative
variations in this sequence can give a specific information concerning
the solar activity.

For the first time, in 1848 a Swiss astronomer R.Wolf proposed
that a measure of the solar activity can be characterised by
the number of sunspots. To this end he recommended to consider
the union of the total number of spots visible on the Sun and
tenfold numbers of regions in which these spots are placed.
This last summand should coordinate the results of
measurements performed under several conditions. Thus, since the
year 1849 the results of daily measurements have become available.
Using more earlier observations and various sources, R.Wolf has
reconstructed the data of the solar activity (with an admissible
accuracy and negligible gaps) up to the year 1818. Now the averaged
number of the sunspots is called the Wolf number.

Later the average monthly values of the Wolf numbers up to the year
1749 (namely this series is used in the present paper) and their
average annual values up to the year 1700 have been reconstructed.
In the last case however, the error can reach more than 10 percent.
The chosen data covers the wide time interval without gaps and a
quite high time resolution. The investigated sequence is shown in
Fig.1. The time with the interval of one month is plotted on the
abscissa. The corresponding Wolf numbers are laid as ordinates.

On the first step of the application of SSA it is necessary
to take the maximum permissible value of parameter $\tau$. For our
investigations we used $\tau=500$. This value allows us to make up
periodicities up to period of 42 years. The use of the larger values
leads to essential computer problems. Moreover, increasing the
parameter $\tau$ (up to 600) does not yield any significant change
in the results of the first expansion in principal components.

Owing to a large value of $\tau$ the dependence of the roots of
eigenvalues of the covariance matrix (ordered in decreasing)
decays exponentially. In addition, the sum of the leading five
eigenvalues exceeds $99\%$ of their sum total. Coupled with
a quite large numbers of initial points, this leads to the fact that
the first principal component yields a small enough smoothing of the
initial series, and by means of the leading four-five components
one can reconstruct this series. Moreover, at small values of
the parameter $\tau$, say at $\tau=5$, the first component weakly
changes its form. This is due to the fact that the SSA method
is stable with respect to this parameter. Therefore, application of
a sufficiently large $\tau$ is justified only for the prediction.

To check possibilities of the prediction let us cut off the initial
sequence of the {\it averaged monthly} values for a length of 216
points (18 years) and reconstruct it by the following way. Evaluate
optimal parameter values of the reconstruction algorithm by means
of an additional truncation and resolve the obtained series into $\tau$
components. Therewith it is necessary to choose such a number of
the leading components at which we get the best coincidence of the
reconstructed values with the additionally truncated data. Then using
the obtained parameters, we will find the initially shortened part of
216 points.

By means of the direct exhaustion one can define that the best results
we get at $r=150$ (the number of the chosen components). Once
again let us take the reduced sequence (for a length of 18 years) and
apply the chosen $r$ for the prediction. One can further improve the
quality of the prediction if we will decompose the predicted interval
(216 points) into subintervals and after prediction of each part
calculate the principal components. In the ideal case, after the
prediction of each point it is necessary to realise this process.
However, such an approach leads to an increase of the computer
time.

Fig.2. illustrates the reconstructed series at which recalculations
have been realised every 72 points (i.e. 3 times). However, as it turns
out, this result is almost identical to the prediction of the whole interval
of 216 points, without decomposition into subintervals.

Theoretically, one can analyse the series components with the
aim of detection the existence of the known and any other periods.
However a number of these periods and thus, the resemblance of
the components $Y_i$ of the initial series makes this analysis a
quite cumbersome procedure. In addition, selection of sufficiently
short periodicities (about several months) is a very difficult computer
task. For these reasons, the choosing a large enough subinterval of
time is the more simply problem.

Let us consider now a series with the {\it average annual} Wolf
numbers. This series contains only 248 points. Thus, the maximum
allowable value of $\tau$ is 123. Choose this value as an initial one.
The corresponding eigenvalues is shown in Fig.3. The first number
presents the principal component related to a trend. The following
step-like data form the pairs of components with the numbers 2--3,
4--5, 6--7, 8--9 and 11--12. Beginning with 14-th number this
dependence gives way to an exponential tail. Eigenvectors for the
pairs 2--3, 4--5, 8--9 and 11--12 (see Fig.4) fit periodicities with
11-year period. The corresponding helical dependence
for the components 2 and 3 is shown in the left side of Fig.5.

It should be noted that except for the evident the eleven-year
solar cycle it is possible to guess a supposed eighty-year Gleisberg
cycle (see Fig.5). Here we keep in mind the pair of the eigenvectors
6 and 7. The corresponding eigenvalues is not exactly equal to
each other, i.e. the step-like data is oblique (see Fig.3) and the phase
lag is not $\pi/2$. That is the reason why the form of the function
diagram is not helical. In spite of this fact and a small enough
eigenvalues, it is quite possible to trace a periodicity.

For the best selection of the eighty-year dynamics one can identify
parameter $\tau$. By the numerical analysis we have found that
$\tau=80$ is the most suitable value. In this case, vectors 4 and 7
are associated with such a periodicity. The diagram for 4-th and 7-th
components is shown in Fig.6. A possible eighty-year solar cycle
obtained by the reconstruction only via these two components is
shown in Fig.7.

Consider now a much more interesting problem concerning the
possibility of the prediction of the average annual Wolf numbers
by the SSA method. Shorten this series from the right hand side
for a length of 18 points (that means years) and reconstruct it.
To make this let us truncate additionally an interval of 11 points
and try to restore it in the best way. Choosing the suited numbers
of eigenvectors we will use such a procedure in various parts of the
series.

For the series of 219 points ($219=248-18-11$, where 248
is the number of all points, see above) the maximal possible
$\tau=109$. As follows from numerical analysis, prediction for
11 points strongly depends on the picking of the components.
At the same time, in a sufficiently wide range a {\it qualitative}
guide of the prediction is not bad. However, for the quantitative
prediction the given $\tau$ is too large. For the lesser values of
$\tau$ it is quite possible to find the necessary number of
components. For example, for $\tau=33$ and choosing the
leading 11-th components we get fairly good results of the
prediction.

Thus, let us use these parameters for the prediction of the removed
18 points. In contrast to the case of the average monthly data
and $\tau=500$ now, at $\tau=33$ we have a possibility to recalculate
eigenvectors with consideration of the last predicted point. This
recalculations can be made at the stage of the selection of vectors
as well as the prediction. The result of the prediction with the
correction at each step is shown in Fig.8.

In addition, we have studied a sequence of natural logarithms
obtained from the initial series. Taking the logarithm is often used
at the data processing (for example, in the case of the correlation
analysis) and allows to get more interesting and complete results.
However, eigenvalues and eigenvectors after such an operation have
not principally changed. Thus, in this case taking the logarithm
is not necessary because the basic information can be obtained
from the analysis of the initial series.

In the closing stage let us consider application of SSA
for a {\it real prediction} of the solar activity. For this purpose,
the average annual sequence of the Wolf numbers from the year 1748
to the year 1996 has been chosen. The end of this series corresponds
to a minimum of the solar activity. Therefore it is interesting to
describe the further activity of the Sun and predict its two next
maxima. To realise this idea it is necessary to extend the average
annual sequence for a length of 18 year points.

We resolved the series into 33 components and choose for the
prediction the leading eleven of them. The result of the prediction is
shown in Fig.9. As follows from this figure, in the nearest future,
in comparison with the two previous maxima, the Sun will be
in a relatively quiescent state. In addition, the level of the forthcoming
maximum in the year 2011 will not be so high. On the basis of
our numerical investigations concerning the application of SSA,
the prediction during more than one and a half of 11-th
year solar cycle is not justified. Nevertheless, we believe that the
obtained values of the Wolf numbers shown in Fig.9 are a quite
reliable.

For comparison, a series of the average monthly Wolf numbers
has been analysed. However, its investigation did not yield
an essential modifications to the obtained results.

\section{Conclusion}

It seems to be true that in the nearest future investigations of time
series by means of SSA will occupy a more important
and deserving place among different ways related to the processing
and forecasting many experimentally obtained sequences. Resolving the
initial series into components the analytical form of which is not
fixed, this method allows us on a quite high level to reveal periodic
subsequences and forecast the dynamics of this series. Therewith,
restrictions to the numbers of points and characteristic periods in
the investigated data as a rule, is essentially less than at
applications of the other methods (for example, at correlation or
Fourier analyses).

In the present article a possibility of the SSA application to the
sequence of the Wolf numbers characterising the solar activity,
is considered. In spite of a relatively small length of this sequence,
SSA allows to reveal the components corresponding
the known solar cycles and, on the basis of only certain constituents,
admits reconstruction of the data. Moreover, it is found that by
means of SSA it is possible to extend short time series with
an acceptable accuracy.

Like any other method, SSA is not devoid of drawbacks.
First, there is a certain difficulty in the problem of revealing
unknown frequencies in the investigated sequence. These
frequencies can be more easily obtained by the Fourier analysis.
Secondly, SSA does not contain clear rules concerning
the choice of components, especially in the case of forecasting.
Finally, SSA does not give the correct prediction of the
cycle period. Therefore in the predicted sequence a systematic
phase lag is accumulated. Nevertheless, as follows from the
performed analysis it is a useful supplement
to the existing methods of the experimental data processing.

In the forthcoming investigations devoted to the analysis of the
solar activity, we will consider a possibility of a certain correction
of the cycle phase by means of additional methods and via special
feature of the sunspot dynamics. Besides, to resolve this problem one
can use an empirical dependence of the amplitude and the phase
of the solar cycle \cite{Dmikuzob}. Moreover, it seems to be
reasonable that it is quite possible to improve control parameters
which are used in SSA.

On the whole we can say that the described method of singular
spectrum analysis is a sufficiently advantageous and promising way
for the prediction of dynamics of the solar activity.

\end{multicols}

\newpage
\vspace*{3cm}

\begin{center}
\begin{figure}[htbp]
\centering
\includegraphics*[bbllx=0.19in,bblly=0.18in,bburx=5.18in,bbury=2.84in,scale=1.12]{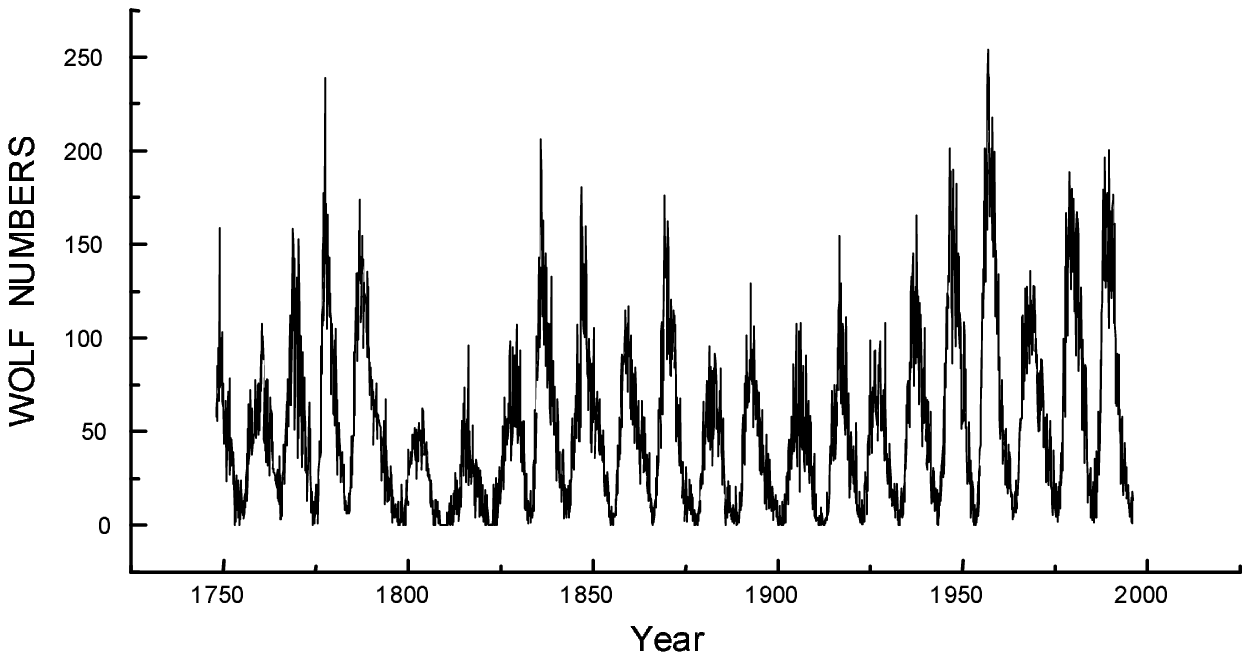}
\caption{ The data of the average monthly values of the
Wolf numbers.}
\end{figure}

\end{center}

\newpage
\vspace*{3cm}

\begin{center}
\begin{figure}[htbp]
\centering
\includegraphics*[bbllx=0.19in,bblly=0.18in,bburx=5.61in,bbury=4.21in,scale=0.74]{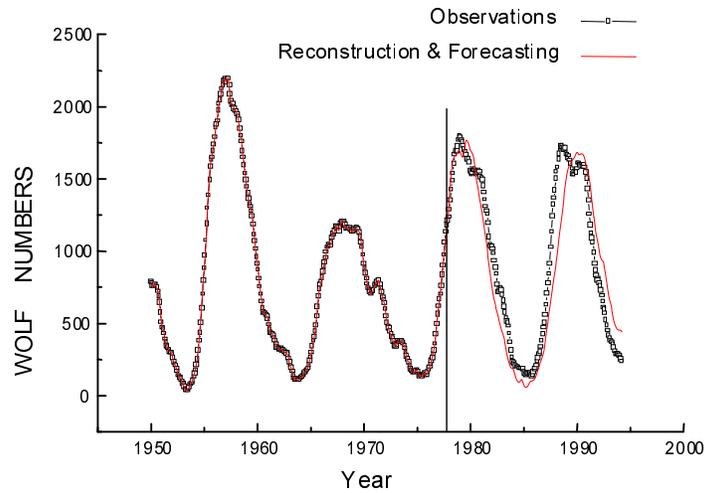}
\caption{ An example of the prediction of the solar activity
for a length of 216 points (18 years) by the average monthly
Wolf numbers. The vertical line corresponds to the boundary
value of the removed points. At the expansion, 500 components
have been applied; in the reconstruction procedure 150
components have been involved. Numerical analysis has been
performed by three stage: after prediction of the next 72 points
recalculations have been made.}
\end{figure}

\end{center}

\newpage
\vspace*{3cm}

\begin{center}
\begin{figure}[htbp]
\centering
\includegraphics*[bbllx=0.19in,bblly=0.18in,bburx=5.46in,bbury=3.96in,scale=1.03]{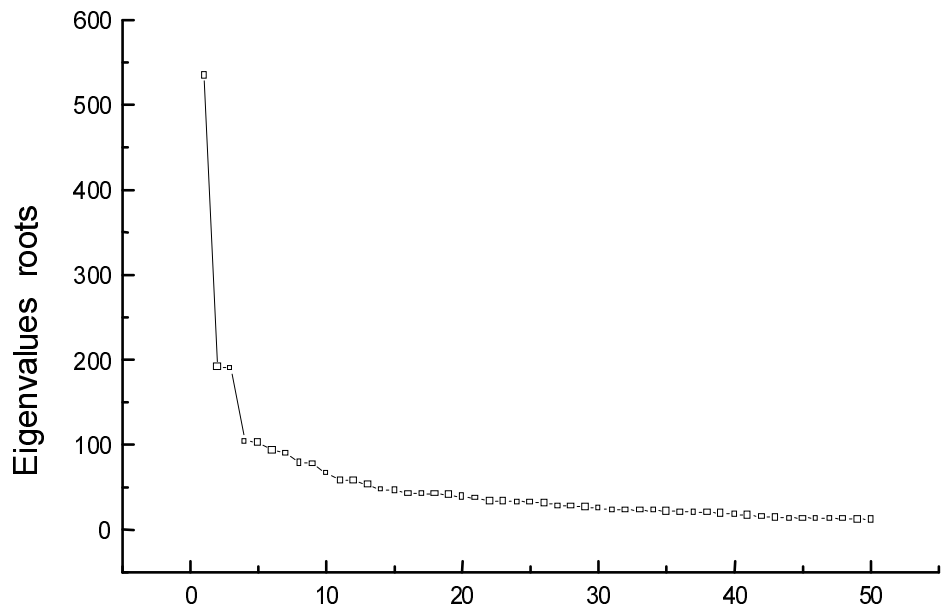}
\caption{ The leading 50 eigenvalues of the covariance matrix
obtained at the expansion of the average annual Wolf numbers
into 123 components.}
\end{figure}

\end{center}

\newpage
\vspace*{3cm}

\begin{center}
\begin{figure}[htbp]
\centering
\includegraphics*[bbllx=0.19in,bblly=0.18in,bburx=6.69in,bbury=3.93in,scale=1.00]{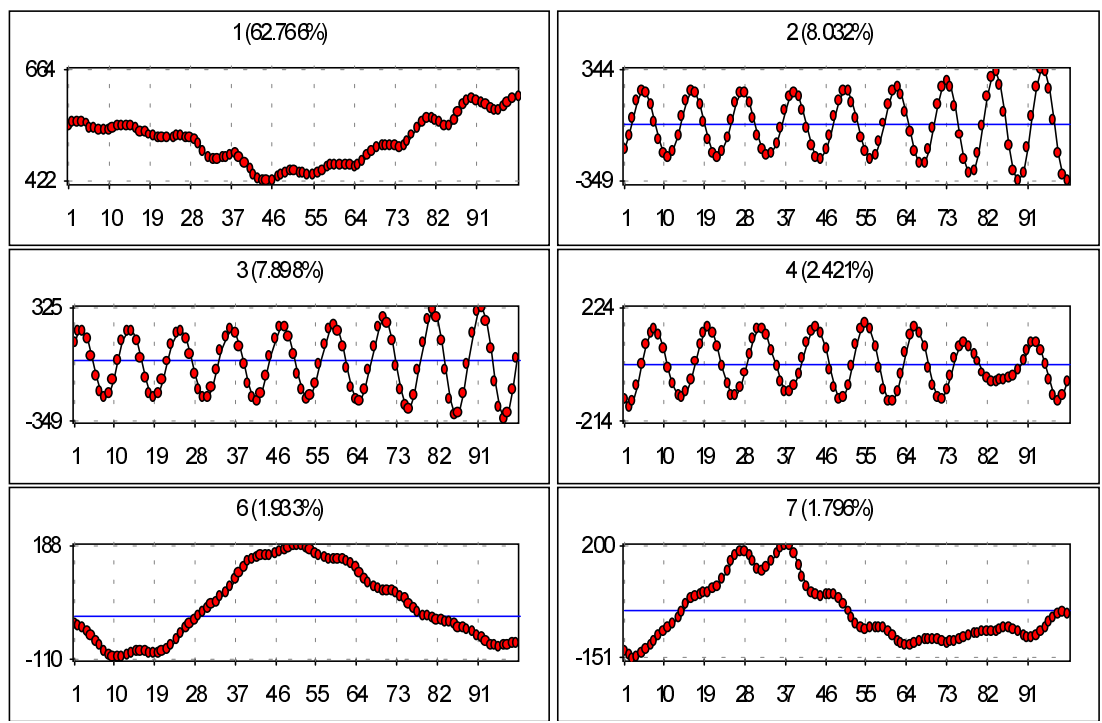}
\caption{ Some of components corresponding to eigenvalues
shown in Fig.3. Their percentage in the initial series is denoted by
the parentheses.}
\end{figure}

\end{center}

\newpage
\vspace*{3cm}

\begin{center}
\begin{figure}[htbp]
\centering
\includegraphics*[bbllx=0.19in,bblly=0.18in,bburx=6.69in,bbury=1.98in,scale=1.00]{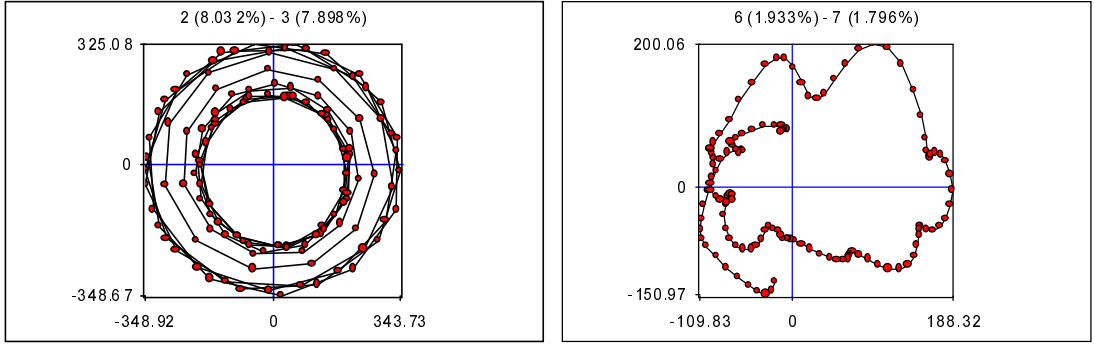}
\caption{ Dependencies of the 2-nd and the 3-rd components (on
the left-hand side) and the 6-th and the 7-th components (on the
right-hand side) shown in Fig.4. The left and the right parts of the
Figure correspond to the 1-st and the 3-rd step-like data of
eigenvalues, respectively.}
\end{figure}

\end{center}

\newpage
\vspace*{3cm}

\begin{center}
\begin{figure}[htbp]
\centering
\includegraphics*[bbllx=0.19in,bblly=0.18in,bburx=6.69in,bbury=2.17in,scale=1.00]{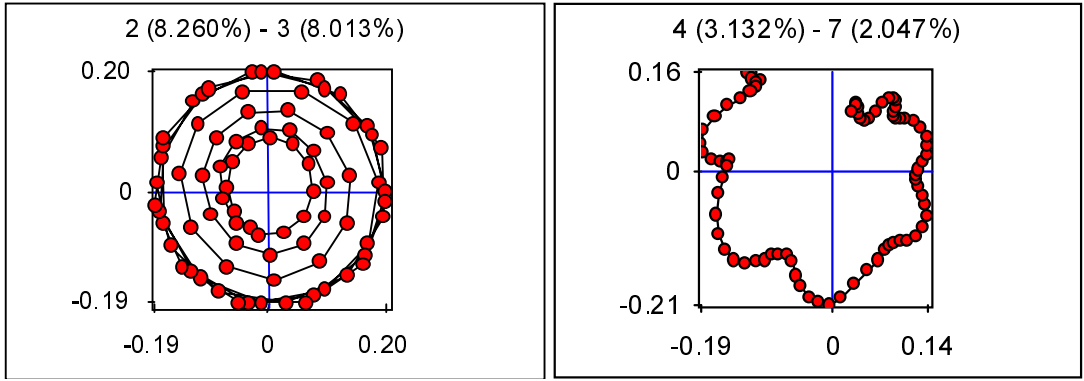}
\caption{ Dependencies of the 2-nd and the 3-rd components (on
the left-hand side) and the 4-th and the 7-th components (on the
right-hand side) at $\tau=80$.}
\end{figure}

\end{center}

\newpage
\vspace*{3cm}

\begin{center}
\begin{figure}[htbp]
\centering
\includegraphics*[bbllx=0.19in,bblly=0.18in,bburx=5.45in,bbury=4.11in,scale=0.76]{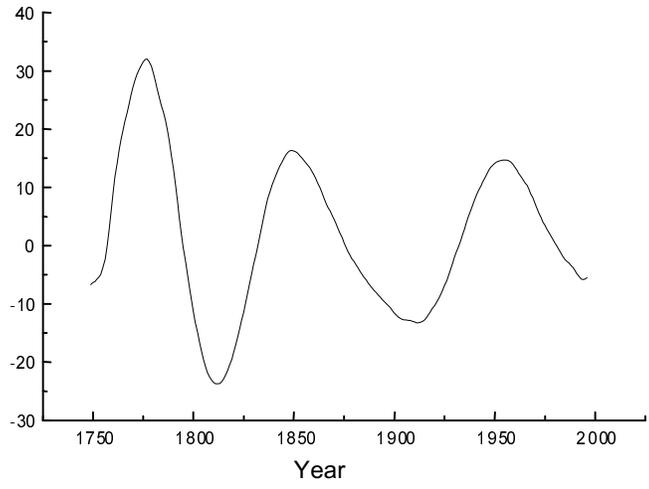}
\caption{ Reconstruction of the 4-th and the 7-th components
corresponding 80-year solar cycle.}
\end{figure}

\end{center}

\newpage
\vspace*{3cm}

\begin{center}
\begin{figure}[htbp]
\centering
\includegraphics*[bbllx=0.19in,bblly=0.18in,bburx=5.57in,bbury=4.21in,scale=1.00]{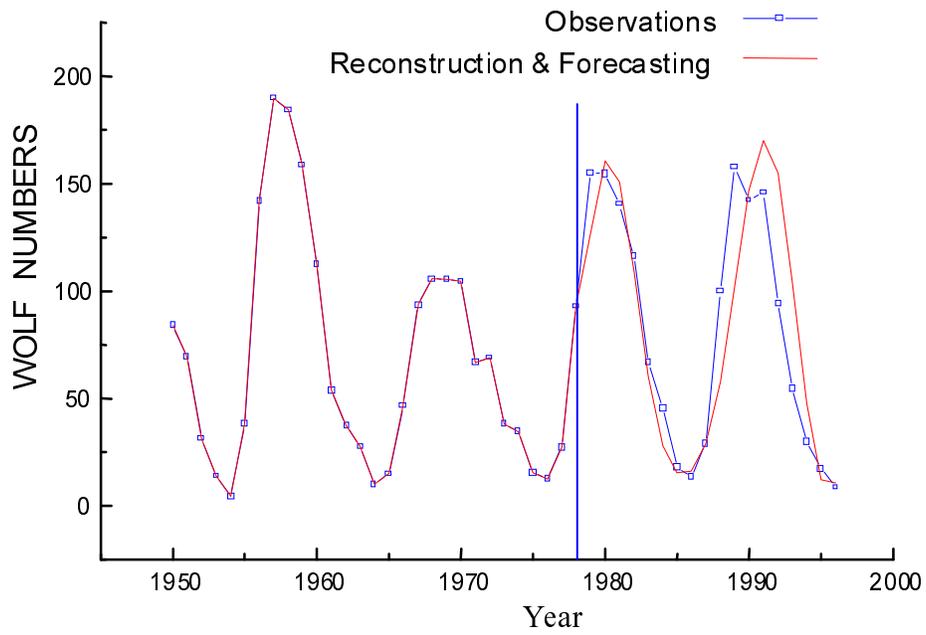}
\caption{ Prediction of the solar activity the each-step correction.
11 in 33 components are chosen. The vertical line marks the
boundary value of the removed 18 points.}
\end{figure}

\end{center}

\newpage
\vspace*{3cm}

\begin{center}

\begin{figure}[htbp]
\centering
\includegraphics*[bbllx=0.19in,bblly=0.18in,bburx=4.46in,bbury=3.36in,scale=1.23]{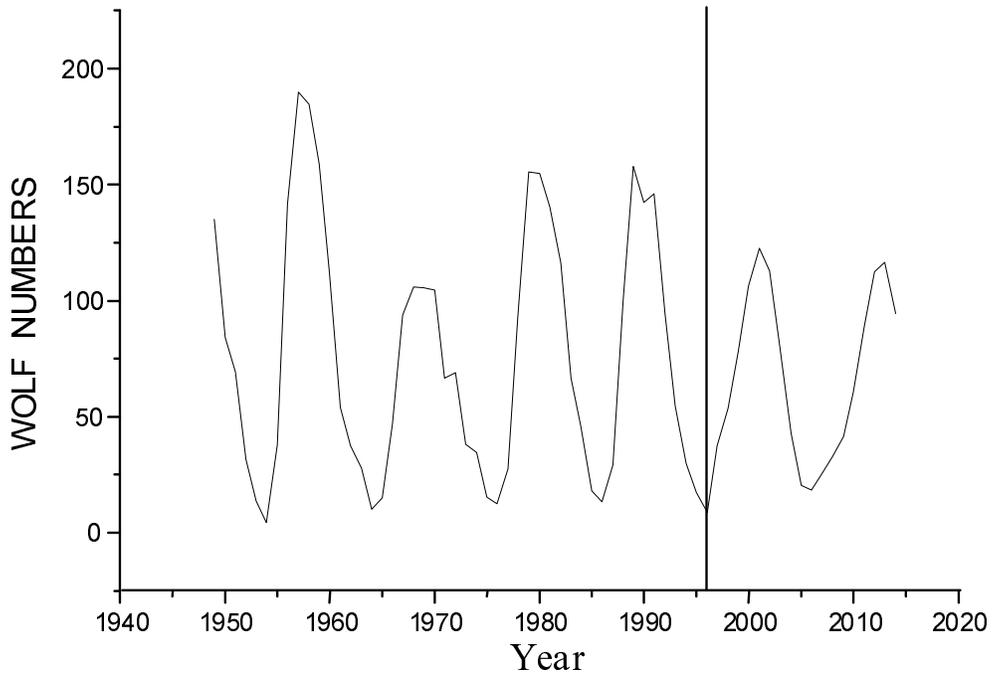}
\caption{ Prediction of the solar activity up to 2015 year.}
\end{figure}

\end{center}

\end{document}